\documentclass[twocolumn,superscriptaddress,secnumarabic,amsmath,amssymb, nobibnotes, aps, prd]{revtex4-1}

\usepackage{amsmath}                    
\usepackage[bbgreekl]{mathbbol}
\DeclareSymbolFontAlphabet{\mathbbl}{bbold}

\usepackage{amsfonts,amssymb,amsthm,mathtools}    
\DeclareSymbolFontAlphabet{\mathbbm}{bbold}
\DeclareSymbolFontAlphabet{\mathbb}{AMSb}%
\usepackage{url}
\usepackage{enumitem}
\usepackage{bm} 
\usepackage[usenames,dvipsnames]{xcolor}
\usepackage{MnSymbol}
\usepackage{hyperref}
\usepackage{tensor}
\usepackage[titletoc,title]{appendix}
\usepackage{mathtools}
\usepackage{xcolor} 
\usepackage{indentfirst} 

\AtBeginDocument{}%

\newcommand{\corurl}{red}
\newcommand{\corcite}{ForestGreen}
\newcommand{\corlink}{blue}

\usepackage{scalerel}

\renewcommand{\d}{\mathrm{d}}
\newcommand{\dd}{\mathchoice
	{\mathbbm{d}\rrule{.087ex}{1.605ex}\hspace*{0.15ex}} 
	{\mathbbm{d}\rrule{.087ex}{1.605ex}\hspace*{0.15ex}} 
	{\mathbbm{d}\rrule{.08ex}{1.125ex}\hspace*{0.15ex}}  
	{\mathbbm{d}\rrule{.06ex}{.8ex}\hspace*{0.15ex}}     
}

\usepackage{graphicx}
\makeatletter
\newcommand\rrule[3][0pt]{%
	\ifdim#2>#3\math@hrule[#1]{#2}{#3}\else\math@vrule[#1]{#2}{#3}\fi}
\newcommand\math@hrule[3][0pt]{%
	\gdef\mystery@factor{0.07}%
	\@tempdima=#3%
	\rule[#1]{0pt}{#3}
	\raisebox{.5\@tempdima+#1}{%
		\makebox[#2][l]{\kern-.5\@tempdima\@@mathrule{#2}{#3}}}%
}
\newcommand\math@vrule[3][0pt]{%
	\gdef\mystery@factor{0.0}%
	\@tempdima=#2%
	\rule[#1]{0pt}{#3}
	\raisebox{-.0\@tempdima+#1}{%
		\kern0.5\@tempdima%
		\rotatebox{90}{\kern-0.5\@tempdima\makebox[#3][l]{\@@mathrule{#3}{#2}}}%
		\kern0.5\@tempdima}%
}
\def\@@mathrule#1#2{%
	\@tempdimb=#2%
	\@tempdima=\dimexpr#1-\mystery@factor\@tempdimb
	\pdfliteral{%
		q []0 d %
		1 J 
		\strip@pt\@tempdimb\space w \strip@pt\@tempdimb\space 0 m %
		\strip@pt\@tempdima\space 0 l S Q }}
\makeatother

\makeatletter
\DeclareFontFamily{OMX}{MnSymbolE}{}
\DeclareSymbolFont{MnLargeSymbols}{OMX}{MnSymbolE}{m}{n}
\SetSymbolFont{MnLargeSymbols}{bold}{OMX}{MnSymbolE}{b}{n}
\DeclareFontShape{OMX}{MnSymbolE}{m}{n}{
<-6>  MnSymbolE5
<6-7>  MnSymbolE6
<7-8>  MnSymbolE7
<8-9>  MnSymbolE8
<9-10> MnSymbolE9
<10-12> MnSymbolE10
<12->   MnSymbolE12
}{}
\DeclareFontShape{OMX}{MnSymbolE}{b}{n}{
<-6>  MnSymbolE-Bold5
<6-7>  MnSymbolE-Bold6
<7-8>  MnSymbolE-Bold7
<8-9>  MnSymbolE-Bold8
<9-10> MnSymbolE-Bold9
<10-12> MnSymbolE-Bold10
<12->   MnSymbolE-Bold12
}{}

\let\llangle\@undefined
\let\rrangle\@undefined
\DeclareMathDelimiter{\llangle}{\mathopen}%
{MnLargeSymbols}{'164}{MnLargeSymbols}{'164}
\DeclareMathDelimiter{\rrangle}{\mathclose}%
{MnLargeSymbols}{'171}{MnLargeSymbols}{'171}
\makeatother

\def\wwedgee{{\setbox0\hbox{\ensuremath{\mathrel{\wedge}}}\rlap{\hbox to \wd0{\hss\hspace*{.6ex}\ensuremath\wedge\hss}}\box0}}

\newcommand{\wwedge}{\mathrel{\wwedgee}}

\newcommand{\et}{{e_{\mathrm{t}}}}
\newcommand{\om}{{\omega_{\mathrm{t}}}}

\hypersetup{linktocpage,colorlinks,urlcolor=\corurl,citecolor=\corcite,linkcolor=\corlink,
pdftitle={A simple symplectic formulation for tetrad gravity},
pdfauthor={J.F. Barbero, B. D\'{\i}az, J. Margalef-Bentabol, E.J.S. Villase\~nor}}

\begin{document}


\title{Concise symplectic formulation for tetrad gravity}

\author{J. Fernando Barbero G.}
\email{fbarbero@iem.cfmac.csic.es}
\affiliation{Instituto de Estructura de la Materia, CSIC. Serrano 123, 28006 Madrid, Spain}
\affiliation{Grupo de Teor\'{\i}as de Campos y F\'{\i}sica Estad\'{\i}stica. Instituto Gregorio Mill\'an (UC3M). Unidad Asociada al Instituto de Estructura de la Materia, CSIC}
\author{Bogar D\'{\i}az}
\email{bdiaz@iem.cfmac.csic.es}
\affiliation{Instituto de Estructura de la Materia, CSIC. Serrano 123, 28006 Madrid, Spain}
\affiliation{Grupo de Teor\'{\i}as de Campos y F\'{\i}sica Estad\'{\i}stica. Instituto Gregorio Mill\'an (UC3M). Unidad Asociada al Instituto de Estructura de la Materia, CSIC}
\affiliation{Departamento de F\'{\i}sica de Altas Energías, Instituto de Ciencias Nucleares, Universidad Nacional Aut\'onoma de M\'exico, Apartado Postal 70-543, Ciudad de M\'exico, 04510, M\'exico}
\author{Juan Margalef-Bentabol}
\email{juanmargalef@psu.edu}
\affiliation{Grupo de Teor\'{\i}as de Campos y F\'{\i}sica Estad\'{\i}stica. Instituto Gregorio Mill\'an (UC3M). Unidad Asociada al Instituto de Estructura de la Materia, CSIC}
\affiliation{Institute for Gravitation and the Cosmos \& Physics Department, Penn State,University Park, PA 16802, USA}
\author{Eduardo J.S. Villase\~nor}
\email{ejsanche@math.uc3m.es}
\affiliation{Grupo de Teor\'{\i}as de Campos y F\'{\i}sica Estad\'{\i}stica. Instituto Gregorio Mill\'an (UC3M). Unidad Asociada al Instituto de Estructura de la Materia, CSIC}
\affiliation{Departamento de Matem\'aticas, Universidad Carlos III de Madrid. Avda.\  de la Universidad 30, 28911 Legan\'es, Spain}



\begin{abstract}
We discuss a simple symplectic formulation for tetrad gravity that leads to the real Ashtekar variables in a direct and transparent way. It also sheds light on the role of the Immirzi parameter and the time gauge.
\end{abstract}

\maketitle



\section{Introduction}

The purpose of this paper is to present a new symplectic formulation for tetrad gravity. Among its most salient features, we would like to highlight the very simple polynomial form of the constraints, its full $SO(1,3)$ invariance and the fact that the Immirzi parameter appears only in the (pre)symplectic form. 

The usual (real) Ashtekar formulation \cite{Ashtekar1,barbero1} can be derived from the results presented here in a straightforward way that illuminates the role of the time gauge. In our opinion, our formulation (which shares some features with the one presented in \cite{Cattaneo}, despite the use of very different methods) provides a viewpoint that neatly complements the one  obtained by using Dirac's algorithm (see, for instance, \cite{Perez,Noui,Montesinos,Montesinos2}). It also sheds light on other issues such as the role of the Immirzi parameter --both at the classical and quantum levels-- and the appearance of constraints quadratic in momenta. 

In general, the Hamiltonian dynamics of a (singular) system is determined by Hamiltonian vector fields $Z$ satisfying
\begin{equation}\label{HVF}
\iota_Z\Omega=\dd H\,,
\end{equation}
where $\Omega$ is a presymplectic form on a phase space $\mathcal{F}$ and $H$ the Hamiltonian of the system. We will denote by $\dd$ and $\wwedge$  the exterior derivative and the wedge product in $\mathcal{F}$, respectively. By requiring the Hamiltonian vector field to be consistent (i.e. tangent to the manifold where the dynamics takes place), the Gotay-Nester-Hinds (GNH) algorithm \cite{GNH1} leads to a sequence of constraint submanifolds  of $\mathcal{F}$. When the algorithm terminates, it provides a constructive and neat method for finding a submanifold  $\Upsilon$ of $\mathcal{F}$ where equation \eqref{HVF} makes sense. 

\section{Symplectic formulation for the Holst Action}
The Hamiltonian description of tetrad gravity discussed here can be obtained from the Holst action \cite{Holst} by using the geometrically inspired GNH method \cite{GNH1,tesis,EM,scalar}. Instead of following this approach, which is interesting in itself and will be presented in an upcoming publication \cite{nos}, we will justify the validity of our formulation by deriving the real Ashtekar formulation from it.

Let $\mathcal{M}$ be a four-dimensional manifold diffeomorphic to $\Sigma\times\mathbb{R}$ where $\Sigma$ is a closed (i.e. compact without boundary), orientable, three-dimensional manifold (this implies that $\Sigma$ is parallelizable). General relativity in tetrad form can be derived from the Holst action \cite{Holst}
\begin{equation}\label{Holst}
S(\bm{\mathrm{e}},\bm{\omega})=\int_{\mathcal{M}}P_{IJKL}\bm{\mathrm{e}}^I\wedge \bm{\mathrm{e}}^J \wedge {\bm{\mathrm{F}}}^{KL}\,,
\end{equation}
where $\bm{\mathrm{e}}^I\in\Omega^1(\mathcal{M})$ are 1-forms (non-degenerate tetrads), ${\bm{\mathrm{F}}}^I_{\phantom{I}\!J}\!:=\mathrm{d}{\bm{\omega}}^I_{\phantom{I}\!J}+{\bm{\omega}}^I_{\phantom{I}\!K}\wedge\, {\bm{\omega}}^K_{\phantom{K}\!J}$ is the curvature of an $SO(1,3)$ connection ${\bm{\omega}}^I_{\phantom{I}\!J}\in\Omega^1(\mathcal{M})$ (we use boldface letters to represent four-dimensional objects),
\[
P_{IJKL}:=\frac{1}{2}\left(\epsilon_{IJKL}+\frac{\varepsilon}{\gamma}\eta_{IK}\eta_{JL}-\frac{\varepsilon}{\gamma}\eta_{JK}\eta_{IL}\right)\,,
\]
$\epsilon_{IJKL}$ is the Levi-Civita symbol ($\epsilon_{0123}=+1$) and $\gamma$ denotes the Immirzi parameter. Here the latin capital indices $I,J\ldots$ range from $0$ to $3$ and are raised and lowered with the help of the Minkowski metric $\eta=(\varepsilon,+1,+1,+1)$ with $\varepsilon=-1$ (we introduce $\varepsilon$ as a simple device to recover the Riemannian case if so desired).

The field equations given by \eqref{Holst} are equivalent to those obtained from the standard Hilbert-Palatini action and can be written as \cite{Holst,Cattaneo} \begin{subequations}
\begin{align}
&\mathrm{D}{\bm{\mathrm{e}}}^I=0\,,\label{Field_eq1}\\
&\epsilon_{IJKL}{\bm{\mathrm{e}}}^J\wedge {\bm{\mathrm{F}}}^{KL}=0\,,\label{Field_eq2}
\end{align}
\end{subequations}
where $\mathrm{D}{\bm{\mathrm{e}}}^I:=\mathrm{d}{\bm{\mathrm{e}}}^I+{\bm{\omega}}^I_{\phantom{I}J}\wedge{\bm{\mathrm{e}}}^J$.

By using the GNH method we arrive at a Hamiltonian formulation defined in a space of fields $\mathcal{F} $ consisting of scalar functions $\et^I,\om^{IJ}\in C^\infty(\Sigma)$ and 1-forms $e^I,\omega^{IJ}\in\Omega^1(\Sigma)$  (we use non-bold fonts for the objects living on $\Sigma$ to distinguish them from those defined on $M$). The basic elements of the formulation are as follows:

\noindent$\bullet$ The field space $\mathcal{F} $ is endowed with the presymplectic form
\begin{align}\label{presymplectic}
\Omega_P=\int_\Sigma \dd\omega^{IJ}\wwedge \dd(P_{IJKL}e^K\wedge e^L)\,.
\end{align}

\noindent$\bullet$  The dynamics is restricted to a submanifold $\Upsilon$ of $\mathcal{F} $ defined by the constraints
\begin{subequations}
\begin{align}
&De^I=0\,,\label{constraint_De}\\
&\epsilon_{IJKL}e^J\wedge F^{KL}=0\,,\label{constraint_eF}
\end{align}
\end{subequations}
where the curvature $F^{I}_{\,\,\,J}:=\mathrm{d}\omega^I_{\,\,\,J}+\omega^I_{\,\,\,K}\wedge\omega^K_{\,\,\,\,\,J}$ satisfies the identity $DF^I_{\,\,\,J}=0$ with $D$ given by a suitable extension of $D\alpha^I=\mathrm{d}\alpha^I+\omega^I_{\,\,\,J}\wedge\alpha^J$ for 1-forms.

\noindent$\bullet$ Let $Z$ be the Hamiltonian vector field tangent to $\Upsilon$ that defines the evolution of the system, then its components satisfy
\begin{subequations}
\begin{align}
&Z_e^I=D\et^I-\omega_{\mathrm{t}\phantom{I}J}^{\phantom{\mathrm{t}}I}e^J\,,\label{ZeI}\\
&\epsilon_{IJKL}\Big(e^J\wedge\big(Z_\omega^{KL}-D\om^{KL}\big)-\et ^J F^{KL}\Big)=0\,,\label{Zomega}\\
&Z_{e\mathrm{t}}^I\,,\,\,\,\,\quad\quad\text{arbitrary}\,,\label{Zebarra}\\
&Z_{\omega\mathrm{t}}^{IJ}\,,\,\,\quad\quad\text{arbitrary}\,.\label{Zomegabarra}
\end{align}
\end{subequations}
On $\mathcal{F}$, the vector field $Z$ is Hamiltonian in the sense that it satisfies $\iota_Z\Omega_P=\dd H$ with
\[
H=\int_\Sigma P_{IJKL}\Big(e^I\!\wedge e^J\!\wedge D\om^{KL}-2\,\et ^I e^J\!\wedge F^{KL}\Big)\,.
\]
Notice that the arbitrariness of $Z_{e\mathrm{t}}^I$ and $Z_{\omega\mathrm{t}}^{IJ}$ implies that $e^I_{\mathrm{t}}$ and $\omega_{\mathrm{t}}^{IJ}$ are themselves arbitrary. This is to be expected as $e_{\mathrm{t}}^I$ play the role of the lapse and the shift, while $\omega_{\mathrm{t}}^{IJ}$ parametrize local Lorentz transformations.

Although the most efficient way to get the previous formulation is to use the GNH method, it can also be obtained by employing the geometric implementation of Dirac's algorithm \cite{Diracnos, HKnos}.

One striking feature of the constraints \eqref{constraint_De} and \eqref{constraint_eF} is their structural resemblance with the field equations \eqref{Field_eq1} and \eqref{Field_eq2}. This suggests a direct approach to obtain the Hamiltonian formulation presented here that takes advantage of the fact that the Holst action is first order, background independent and it is written in terms of differential forms. Actually, there is a very quick and neat way to get equations \eqref{constraint_De} to \eqref{Zomegabarra} as \emph{necessary} conditions. This is a consequence of the fact that differential forms, pullbacks, and the exterior derivative interact in a natural way.  Although in order to prove that they are sufficient some additional work is necessary (tangency requirements must be checked) it is very useful to know that there is a simple way to write the constraints (a fact that is not obvious at all within Dirac's approach).

The starting point is the field equations \eqref{Field_eq1} and \eqref{Field_eq2} which are equivalent to those given by  the Holst action. Let us introduce on $\mathcal{M}$ a foliation defined by the level surfaces $\Sigma_\tau$ of a scalar function $\tau$, a vector field $\partial_\tau\in\mathfrak{X}(\mathcal{M})$ transverse to the foliation with $\d\tau(\partial_\tau)=1$, and the inclusion $\jmath_\tau:\Sigma_\tau\hookrightarrow M$. Finally, let us introduce
\begin{align*}
&{\bm{\mathrm{e}}}_{\mathrm{t}}^I:=\iota_{\partial_\tau}{\bm{\mathrm{e}}}^I\in C^\infty(\mathcal{M})\,, \\
&\underline{\bm{\mathrm{e}}}^I:={\bm{\mathrm{e}}}^I-\mathrm{d}\tau\wedge {\bm{\mathrm{e}}}_{\mathrm{t}}^I\in\Omega^1(\mathcal{M})\,,\\
&{\bm{\omega}}_{\mathrm{t}}^{IJ}:=\iota_{\partial_\tau}{\bm{\omega}}^{IJ}\in C^\infty(\mathcal{M})\,,\\
&\underline{\bm{\omega}}^{IJ}:={\bm{\omega}}^{IJ}-\mathrm{d}\tau\wedge {\bm{\omega}}_{\mathrm{t}}^{IJ}\in\Omega^1(\mathcal{M})\,,
\end{align*}
($\iota$ denotes the interior product) so that
\begin{align*}
&{\bm{\mathrm{e}}}^I=\underline{\bm{\mathrm{e}}}^I+\mathrm{d}\tau\wedge {\bm{\mathrm{e}}}_{\mathrm{t}}^I\,,\\
&{\bm{\omega}}^{IJ}=\underline{\bm{\omega}}^{IJ}+\mathrm{d}\tau\wedge {\bm{\omega}}_{\mathrm{t}}^{IJ}\,.
\end{align*}
Now, if we pullback \eqref{Field_eq1} and \eqref{Field_eq2} to $\Sigma_\tau$ and define $\et^I:=\jmath_\tau^*{\bm{\mathrm{e}}}_{\mathrm{t}}^I$, $e^I:=\jmath_\tau^*\underline{\bm{\mathrm{e}}}^I$, $\om^{IJ}:=\jmath_\tau^*{\bm{\omega}}_{\mathrm{t}}^{IJ}$ and $\omega^{IJ}:=\jmath_\tau^*\underline{\bm{\omega}}^{IJ}$, we get \eqref{constraint_De} and \eqref{constraint_eF}. If we take the interior product of \eqref{Field_eq1} and \eqref{Field_eq2} with $\partial_\tau$ and then pullback the result to $\Sigma_\tau$ we obtain \eqref{ZeI}-\eqref{Zomegabarra}.

\section{Deriving the real Ashtekar formulation}
We derive now the real Ashtekar formulation from the symplectic description given above. This can be taken as an independent consistency check of our formulation.

The presymplectic form \eqref{presymplectic} can be written as (here, $\epsilon_{ijk}$ is the three-dimensional Levi-Civita tensor with $\epsilon_{123}=+1$)
\begin{align}\label{presymplectic_5}
\Omega_P=&\int_\Sigma\dd\left(\omega_{ij}+\varepsilon\gamma \epsilon_{ijk}\omega^{0k}\right) \wwedge \dd\left(\frac{\varepsilon}{\gamma}e^i\wedge e^j\right)\nonumber\\
&+\int_\Sigma\dd\left(\frac{2}{\gamma}\omega^{0}_{\phantom{0}i}+\epsilon_i^{\phantom{i}jk}\omega_{jk}\right) \wwedge \dd(e^0\wedge e^i)\,,
\end{align}
by considering the different terms with $I,J\ldots=0$ and $I,J\ldots=i\,,j\ldots$ ($i,j=1,2,3$). The particular form of \eqref{presymplectic_5} hints at the possibility of finding canonically conjugate variables. Notice, however, that something does not quite fit. On one hand, the $2$-forms $e^i\wedge e^j$ have nine independent components written in terms of the nine independent components of $e^i$. On the other hand, a direct counting shows that $\omega_{ij}+\varepsilon\gamma \epsilon_{ijk}\omega^{0k}$ consists of three 1-forms labeled by the antisymmetrized pair $ij$ (nine independent components) but it is written in terms of \emph{eighteen} independent objects (the components of $\omega^{0k}$ and $\omega_{ij}$). Similar considerations apply to the second integral in \eqref{presymplectic_5}.

In order to find \emph{bona fide} canonically conjugate variables and solve this apparent mismatch, we consider a partial gauge fixing (time gauge), $e^0=0$, and pullback all geometric objects to the submanifold $\mathcal{F}_0:=\{e^0=0\}\subset\mathcal{F}$ given by this gauge condition. In particular, this fixes some of the arbitrary pieces of the components of the Hamiltonian vector field. Specifically, we must have 
\begin{equation}\label{cond}
0=Z^0_e=\d\et ^0+\omega^0_{\phantom{0}i}\et ^i-\omega^{\phantom{\mathrm{t}}0}_{\mathrm{t}\phantom{0}i}\,e^i\,.
\end{equation}
As $\et ^0$ and $\et ^i$ will play in the following the role of lapse and shift, the best course of action is to solve \eqref{cond} for $\omega^{\phantom{\mathrm{t}}0}_{\mathrm{t}\phantom{0}\,i}$. By doing this, the three components of $\omega^{\phantom{\mathrm{t}}0}_{\mathrm{t}\phantom{0}\,i}$ are fixed and the boost part of the $SO(1,3)$ symmetry of the Holst action is broken.

The pull-back of the symplectic form $\Omega_P$ to $\mathcal{F}_0$ is obtained just by plugging $e^0=0$ into \eqref{presymplectic_5}
\begin{align}\label{presymplectic_3}
\Omega_0=\int_\Sigma\dd\left(\omega_{ij}+\varepsilon\gamma \epsilon_{ijk}\omega^{0k}\right)\wwedge\dd\left(\frac{\varepsilon}{\gamma}e^i\wedge e^j\right)\,.
\end{align}

We discuss now in detail the constraints in the time gauge.\vspace*{1ex}

\noindent$\blacktriangleright\ De^I=0$ for $I=i$,
\begin{align}
0&=\mathrm{d}e^i+\omega^i_{\phantom{i}0}\wedge e^0+\omega^i_{\phantom{i}j}\wedge e^j  \nonumber\\
&\overset{\text{time gauge}}{\longrightarrow}\qquad\mathrm{d}e^i+\omega^i_{\phantom{i}j}\wedge e^j=0\,.\label{Deitg}
\intertext{$\blacktriangleright\ De^I=0$ for $I=0$,}
\label{De0tg}
0&=\mathrm{d}e^0+\omega^0_{\phantom{0}i}\wedge e^i  \nonumber\\
&\overset{\text{time gauge}}{\longrightarrow}\qquad\omega^0_{\phantom{0}i}\wedge e^i=0 \,.
\end{align}
The key insight to arrive at the Ashtekar formulation for arbitrary values of the Immirzi parameter is to solve for $\omega^i_{\phantom{i}j}$ in \eqref{Deitg}. We do this by writing $\omega^i_{\phantom{i}j}=-\epsilon^i_{\phantom{i}jk}\Gamma^k$ and solving
\begin{align}\label{eq_gamma}
\mathrm{d}e^i+\epsilon^i_{\phantom{i}jk} \Gamma^j\wedge e^k=0\,.
\end{align}

Plugging the expression for $\Gamma^i$ in \eqref{presymplectic_3} leads to
\begin{align}\label{presymplectic_6}
\Omega_0=\!\!\int_\Sigma\!\! \dd\left(\Gamma^i\!-\varepsilon\gamma \omega^{0i}\right)\!\wwedge\dd\left(-\frac{\varepsilon}{\gamma}\epsilon_{ijk}e^j\wedge e^k\right),
\end{align}
so we can define the following pair of canonically conjugate variables:
\begin{subequations}
\begin{align}
&A^i:=\Gamma^i-\varepsilon\gamma \omega^{0i}\,,\label{Ashtekar_connection}\\
&E_i:=-\frac{\varepsilon}{\gamma}\epsilon_{ijk}e^j\wedge e^k\,.\label{inverse_triad}
\end{align}
\end{subequations}
Notice that both of them depend on nine independent objects: the components of $\omega^{0i}$ and those of $e^i$, respectively. In the following, we will rewrite the constraints in terms of $A^i$ and $E^i$.

As we have already solved \eqref{Deitg}, the only remaining condition coming from $De^I=0$ is \eqref{De0tg}. This gives the usual Gauss law of the Ashtekar formulation because
\begin{align*}
\omega^0_{\phantom{0}i}\wedge e^i=0&\Leftrightarrow (A_i-\Gamma_i)\wedge e^i=0 \\
&\Leftrightarrow \mathrm{d}E^i+\epsilon^{ijk}A_j\wedge E_k=0\,.
\end{align*}
This can be written in terms of the vector density $\tilde{E}^i$ associated with the 2-form $E^i$ in the usual way
\begin{align}\label{Gauss}
\mathrm{div}_w\tilde{E}_i+\epsilon_{ijk}\iota_{\tilde{E}^k}A^j=0 \,,
\end{align}
where the volume form $w\in\Omega^3(\Sigma)$ is given by $3! w=\epsilon_{ijk}e^i\wedge e^j\wedge e^k$.\vspace*{1ex}

\noindent$\blacktriangleright\ \epsilon_{IJKL}e^J\wedge F^{KL}=0$ for $I=i$,
\begin{align}
&0=2\epsilon_{ijk}e^j\wedge F^{0k}-\epsilon_{ijk}e^0\wedge F^{jk}  \nonumber\\
&\overset{\text{time gauge}}{\longrightarrow}\ \ \epsilon_{ijk} e^j\wedge\left(\mathrm{d}\omega^{0k}+\omega^{0}_{\phantom{0}l}\wedge \omega^{lk}\right)=0\,.
\end{align} 
Taking into account that $\varepsilon\gamma\omega^{0k}=\Gamma^k-A^k$ and $\omega^i_{\phantom{i}j}=-\epsilon^i_{\phantom{i}jk}\Gamma^k$, the previous condition becomes
\begin{align*}
\epsilon_{ijk} \Big(&e^j\wedge \mathrm{d}(A^k-\Gamma^k)  \\
&-
\epsilon^{lk}_{\phantom{lk}m}e^j\wedge (A_l-\Gamma_l)\wedge \Gamma^m \Big)=0\,,
\end{align*}
which can be rewritten in the form
\begin{align*}
&\epsilon_{ijk}e^j\wedge(F^k-R^k) \\
&\qquad + e^j\wedge(A_j-\Gamma_j)\wedge(A_i-\Gamma_i)=0\,,
\end{align*}
where 
\begin{align*}
    F^i&:=\mathrm{d} A^i+\frac{1}{2} \epsilon^i_{\phantom{i}jk}A^j \wedge A^k \,,\\ R^i&:=\mathrm{d} \Gamma^i+\frac{1}{2}\epsilon^i_{\phantom{i}jk} \Gamma^j \wedge \Gamma^k.
\end{align*}
 Now, using the Gauss law and the identity $\epsilon_{ijk} e^j \wedge R^k=0$, we finally get
\begin{align}\label{vectorconstraintAshtekar}
\epsilon_{ijk}e^j\wedge F^k=0\,.
\end{align}
In terms of the density $\tilde{E}^i$ this expression takes the usual form of the vector constraint
\begin{align}\label{vectorconstraintAshtekar2}
\iota_{\tilde{E}_i}  F^i=0 \,.
\end{align}

\noindent$\blacktriangleright\ \epsilon_{IJKL}e^J\wedge F^{KL}=0$ for $I=0$,
\begin{align*}
\epsilon_{ijk}e^i\wedge(\mathrm{d}\omega^{jk}+\omega^{j}_{\phantom{j}0}\wedge\omega^{0k}+\omega^{j}_{\phantom{j}l}\wedge\omega^{lk})=0\,.
\end{align*}
Using again $\varepsilon\gamma\omega^{0k}=\Gamma^k-A^k$ and $\omega^i_{\phantom{i}j}=-\epsilon^i_{\phantom{i}jk}\Gamma^k$, we can rewrite the previous expression as
\begin{align}\label{esc_1}
2e^i\wedge R_i+\frac{\varepsilon}{\gamma^2}\epsilon_{ijk}e^i \wedge \left(A^j-\Gamma^j \right) \wedge \left(A^k-\Gamma^k \right)=0\,.
\end{align}
By computing the exterior derivative of \eqref{De0tg} and using \eqref{eq_gamma}, we can write
\[2 e_i \wedge \left( F^i-R^i \right)-\epsilon_{ijk}e^i \wedge \left( A^j-\Gamma^j \right) \wedge \left( A^k-\Gamma^k \right)=0\,.
\]
Plugging this into \eqref{esc_1}, we obtain
\begin{align}
2&e_i \wedge F^i\\
&-\left( 1-\frac{\varepsilon}{\gamma^2}\right)\epsilon_{ijk}e^i \wedge \left( A^j-\Gamma^j \right) \wedge \left( A^k-\Gamma^k \right)=0\,,\nonumber
\end{align}
or equivalently,
\begin{align*}
e_i \wedge \left( F^i+\left(\varepsilon\gamma^2-1 \right) R^i \right)=0\,,
\end{align*}
which, in terms of $\tilde{E}^i$ becomes the familiar scalar constraint of the real Ashtekar formulation
\begin{align}\label{esc_3}
\epsilon_{ijk} \imath_{\tilde{E}_i} \imath_{\tilde{E}_j} \left( F^k+ \left(\varepsilon\gamma^2-1 \right) R^k \right)=0 \,.
\end{align}

\section{Conclusions}

\noindent We end the paper with several comments.

\begin{itemize}
\item[i)] The fact that we have been able to obtain the real Ashtekar formulation for general relativity provides a proof \emph{a posteriori} of the soundness of our approach (which, we emphasize again, can be obtained from the Holst action).
\item[ii)] It is important to point out that \eqref{Zebarra} and \eqref{Zomegabarra} tell us that at every instant of time $\et ^0$ and $\et ^i$ can be taken to be arbitrary. This allows us to remove them from the list of configuration variables of the system and just think of them as given functions of time. These objects are the lapse $N:=\et ^0$, the shift $N^i:=\et ^i$, and $\omega_{\mathrm{t}}^{IJ}$ are the parameters of the local Lorentz transformations.
\item[iii)] At variance with the situation with the presymplectic form \eqref{presymplectic} on $\mathcal{F} $, the final symplectic form \eqref{presymplectic_6} is \emph{independent} of the Immirzi parameter $\gamma$ because the term involving $\Gamma^i$ is actually zero (remember that $\Sigma$ is closed). There is nothing strange here because we know that \eqref{Ashtekar_connection} defines a canonical transformation. The symplectic form is also independent of $\gamma$ when written in terms of the new variables $A^i$ and $E_i$ but then the Hamiltonian constraint becomes $\gamma$-dependent.
\item[iv)] By removing the $\Gamma^i$ term from \eqref{presymplectic_6}, it is straightforward to get the $SO(1,3)$-ADM formulation by using the canonical variables $K^i:=\omega^{0i}$ and $E_i:=\epsilon_{ijk}e^j\wedge e^k$. 
\item[v)] The role of the usual quadratic constraints in momenta \cite{Ashtekar:1991hf} is also clarified in our approach. As it can be seen, by using the time gauge $e^0=0$ and pulling back to the submanifold $\mathcal{F}_0:=\{e^0=0\}\subset\mathcal{F}$, we end up with a well defined symplectic structure --in canonical form-- on the phase space defined by the Ashtekar variables. The counting issues that lead to the introduction of quadratic constraints involving momenta simply disappear.
\item[vi)] The formulation presented here is fully $SO(1,3)$ invariant. If we stick to it, the presymplectic form \eqref{presymplectic} depends on $\gamma$, so the Immirzi parameter should play a role at the quantum level. This may also be the case --both at the classical and quantum levels-- if surface terms are added to the Holst action.
\item[vii)] The Hilbert-Palatini action, as well as the corresponding field equations, can be formally recovered by taking the $\gamma\rightarrow \infty$ limit. Notice, however, that the canonically conjugate variables \eqref{Ashtekar_connection} and \eqref{inverse_triad} are not defined in this limit. This explains why the Ashtekar formulation cannot be derived from the Hilbert-Palatini action.
\end{itemize}


%
%
\begin{acknowledgments}
This work has been supported by the Spanish Ministerio de Ciencia Innovaci\'on y Universidades-Agencia Estatal de Investigaci\'on Grant No. FIS2017-84440-C2-2-P. Bogar D\'{\i}az was supported by the Consejo Nacional de Ciencia y Tecnolog\'{\i}a  (M\'exico) postdoctoral research fellowship Grant No. 371778 and, now, with a DGAPA-UNAM postdoctoral fellowship. Juan Margalef-Bentabol is supported by the Eberly Research Funds of Penn State, by the NSF Grant No. PHY-1806356 and by the Urania Stott Fund of the Pittsburgh Foundation, Grant No. UN2017-92945.
\end{acknowledgments}

%
%


\end{document}